\documentclass[prl,apsrev4-1,12pt]{revtex4-1}

\usepackage{amsfonts}
\usepackage{amsmath}
\usepackage{gensymb}
\usepackage{graphicx}
\usepackage{bm}
\usepackage{color}
\usepackage{amssymb}
\usepackage{ulem}
\usepackage{times}
\usepackage{dcolumn}
\usepackage{cases}
\usepackage{txfonts}
\usepackage{textcomp}
\usepackage[mathscr]{euscript}

\begin{document}

\title{Superstructures of Cubic and Hexagonal Diamonds Comprising a Family of Novel $sp^3$ Superhard Carbon Allotropes}
\author{Hui-Juan Cui$^1$, Xian-Lei Sheng$^2$, Qing-Bo Yan$^3$, Qing-Rong Zheng$^{1,*}$, and Gang Su$^{1,*}$ }
\affiliation{\vspace{0.5cm}$^1$Theoretical Condensed Matter Physics and Computational Materials Physics Laboratory, School of Physics, University of Chinese Academy of Sciences, Beijing 100049, China\\
$^2$Beijing National Laboratory for Condensed Matter Physics,
 Institute of Physics, Chinese Academy of Sciences, Beijing 100190, China\\
$^3$College of Materials Science and Opto-Electronic
Technology, University of Chinese Academy of Sciences, Beijing
100049, China \\ \\
$^*$Correspondence and requests for materials should be addressed to \\
GS (email: gsu@ucas.ac.cn) or QRZ (email: qrzheng@ucas.ac.cn).}

\begin{abstract}
 Superstructures of cubic and hexagonal diamonds (h- and c-diamond) comprising a family of stable diamond-like $sp^3$ hybridized novel carbon allotropes are proposed, which are referred to as U$_n$-carbon where $n \geq 2$ denotes the number of structural layers in a unit cell. The conventional h- and c-diamond are included in this family as members with $n=2$ and $3$, respectively. U$_n$-carbon ($n=4-6$), which are unveiled energetically and thermodynamically more stable than h-diamond and possess remarkable kinetic stabilities, are shown to be insulators with indirect gaps of $5.6 \sim 5.8$ eV, densities of $ 3.5 \sim 3.6$ g/cm$^3$, bulk modulus of $4.3 \sim 4.4 \times 10^{2}$ GPa, and Vickers hardness of $92.9 \sim 97.5$ GPa even harder than h- and c-diamond. The simulated x-ray diffraction and Raman spectra are presented for experimental characterization. These new structures of carbon would have a compelling impact in physics, chemistry, materials science and geophysics.

\end{abstract}

\maketitle


Carbon is a common nonmetal element ubiquitous on earth, which  can form $sp$, $sp^{2}$ and $sp^{3}$ hybridized chemical bondings \cite{hybridized}, thereby capable of forming various allotropes such as, typically, graphite, diamond, amorphous carbon, fullerenes \cite{c60}, carbon nanotubes \cite{cnt}, graphene \cite{graphene}, and so on. Diamond has two well-known structures, cubic and hexagonal diamond (c- and h-diamond), which can be formed naturally in the interior of earth or meteorites, or can be synthesized artificially by diverse means in laboratory. Among others, it has been realized that h-diamond can be obtained from graphite at temperature 1000$\celsius$ and pressure $\thicksim$ 13 GPa \cite{hdiamond}, whose simulated hardness on (100) plane is even 58\% larger than c-diamond \cite{press}; the ultrahard polycrystalline cubic diamond can be synthesized by direct conversion of graphite under static high presure and temperature \cite{Irifune}; under cold compression graphite was also found to transform into a new superhard carbon allotrope \cite{superhard}, whose structure is still under debate. On the other hand, diamonds can be utilized to fabricate distinct superlattices, e.g., by producing multilayer structures of isotopically pure $^{12}$C and $^{13}$C \cite{watanabe}, or by using nano-crystalline diamond films \cite{chimowa}. In spite of these carbon structures, however, whether stable diamond superstructures that somehow employ  partial unit structures of c- and h-diamond can exist, is an intriguing but challenging issue, because it will probably not only provide new ways to synthesize diamond-like carbon allotropes with more compelling properties than conventional diamonds, but also might shed new light on the formation mechanism of superhard materials.


Here we propose a general constructing scheme of novel allotropes of carbon, which allows us to generate a family of 3D diamond-like $sp^{3}$ hybridized structures of carbon that we refer to as U$_n$-carbon ($n=2, 3 ,4, 5, \dots$), where $n$ characterizes the number of constructing layers with each position occupied by carbon-carbon (C-C) unit in a unit cell. It is interesting that U$_n$-carbon include h- and c-diamond as its simplest members with $n=2$ and $3$, respectively, and  could also be obtained from graphite at high pressure and temperature. U$_n$-carbon can be viewed as superstructures of c- and h-diamond, as they are formed by stacking the partial unit structures of c- and h-diamond in sequences. By means of the first-principles calculations, we found that U$_n$-carbon ($n=4-6$), which are energetically and thermodynamically even more stable than h-diamond but comparable with c-diamond, are insulators with indirect gaps of $5.6 \sim 5.8$ eV, densities of $ 3.5 \sim 3.6 $ g/cm$^3$, bulk modulus of $4.3 \sim 4.4 \times 10^{2}$ GPa, and Vickers hardness of $92.9 \sim 97.5$ GPa even harder than h- and c-diamond. The simulated x-ray diffraction and Raman spectra useful for further experimental identification are presented. The transforming pathways from graphite to U$_n$-carbon are also addressed.
\bigskip

\begin{figure}[tbp]
\includegraphics[width=0.55\linewidth,clip]{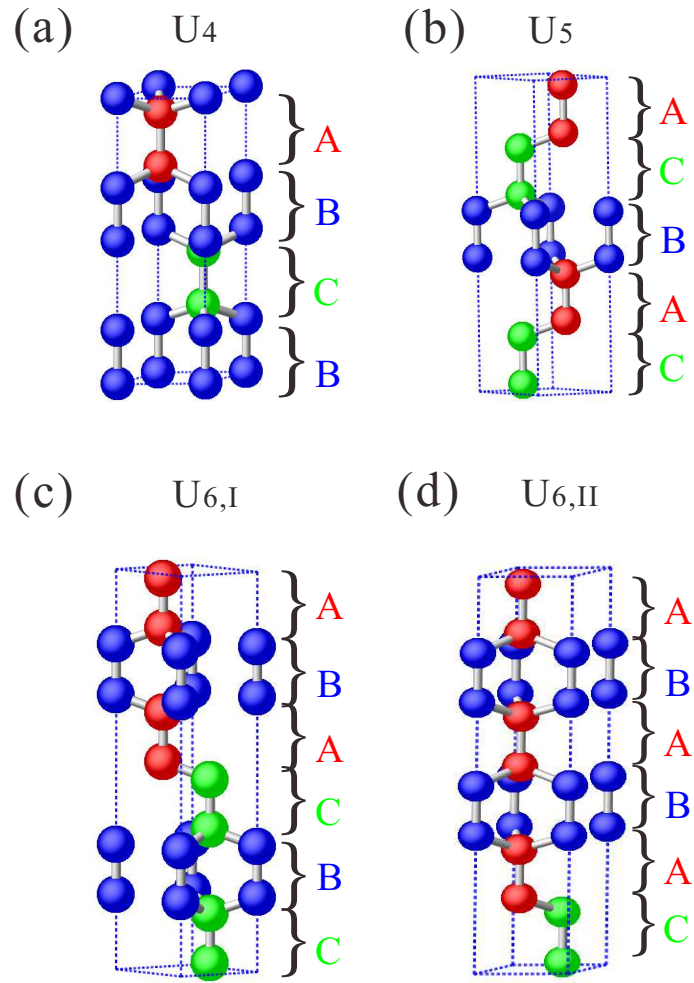}
\caption{The geometrical structures of novel carbon allotropes U$_n$-carbon in a unit cell. (a) U$_4$-carbon ($n=4$, ABCB), (b) U$_5$-carbon ($n=5$, ACBAC), (c) U$_{6,I}$-carbon  ($n=6$, ABACBC), and (d) U$_{6,II}$-carbon ($n=6$, ABABAC). The letters A, B and C indicate the inequivalent positions of C-C unit, which also represent the inequivalent sites in the closed-packed planes from top view.}
\end{figure}

\noindent{\textbf{Results}}

\noindent{\textbf{Structures}}. By carefully analyzing the structural motifs of c- and h-diamond, one may note that they have structural connections with hexagonal closed packed (hcp) and face-centered-cubic (fcc) structures. c-diamond has ABCABC$\cdots$ stacking sequence along [111] direction, while h-diamond has the stacking sequence of ABAB$\cdots$ along [0001] direction. Viewing the arrangement of atoms in c- and h-diamond from the top, one can find that they are the same as those of fcc and hcp structures, respectively (see supplementary information, Figs. S1 and S2). However, this kind of stacking is different from hcp and fcc structures, where the positions A, B and C are not occupied by one carbon atom but by a pair of carbon atoms that occupy the same position. If we introduce more layers comprised of carbon-carbon (C-C) units into c-diamond, there must exist new structures of carbon. According to this scheme, we have generated a family of novel carbon  allotropes that we refer to as U$_n$-carbon. The geometrical structures of U$_4$-, U$_5$-, U$_{6,I}$- and U$_{6,II}$-carbon for $n=4, 5, 6$, respectively, are shown in Fig. 1 as examples. For $n=6$, there are two inequivalent structures that we dub I and II phases. For $n=7$, there are three inequivalent structures, and for $N \geq 8$, more than three inequivalent structures can be obtained. Of the most interesting is that the side views of U$_n$-carbon look quite different, while their top views are the same as that of c-diamond as well as fcc structure (Fig. S2). U$_n$-carbon can be viewed as a kind of superstructures formed by the structural units of h- and c-diamond. For instance, U$_4$-carbon has only one independent closed packed structure ABCB, which is the superstructure comprised of two structural units AB and CB of h-diamond; U$_5$-carbon is the superstructure of the c-diamond unit ABC and h-diamond unit AC; U$_{6,I}$-carbon is the superstructure of two c-diamond structural units ACB and CAB; and U$_{6,II}$-carbon is the superstructure of three structural units AB, AB and AC of h-diamond, as manifested in Fig. 1.
\bigskip

\noindent{\textbf{Physical parameters}}. By means of the first-principles calculations, the geometric structures of U$_4$-, U$_5$-, U$_{6,I}$- and U$_{6,II}$-carbon are fully optimized, and their structural and physical parameters are calculated, as listed in Table I. U$_4$-carbon is built in the stacking sequence of ABCB  [Fig. 1(a)] and has two inequivalent Wyckoff positions $4e(0.000,0.000,0.093)$ and $4f(2/3,1/3,0.844)$. U$_5$-carbon is piled up in the sequence of ABCAC (or ACBAC)  [Fig. 1(b)],
which belongs to the trigonal crystal system and has five inequivalent Wyckoff positions
 $2c(0.000,0.000,0.575)$, $2d(2/3,1/3,0.975)$, $2d(2/3,1/3,0.823)$, $2d(2/3,1/3,0.227)$ and $2d(2/3,1/3,0.376)$. U$_6$-carbon has two different structures, where U$_{6,I}$-carbon is piled up in sequence of ABACBC (or ABCACB) [Fig. 1(c)] and has three inequivalent Wyckoff positions $4e(0.000,0.000,0.687)$, $4f(2/3,1/3,0.844)$ and $4f(2/3,1/3,0.521)$; and U$_{6,II}$-carbon is stacked in the sequence of ABABAC [Fig. 1(d)] and has six inequivalent Wyckoff positions
$2g(0.000,0.000,0.730)$, $2g(0.000,0.000,0.605)$, $2i(2/3,$ $1/3,0.063)$, $2h(1/3,2/3,0.772)$, $2h(1/3,2/3,0.563)$ and $2h(1/3,2/3,0.105)$. All of these structures have two unequal bond lengths $d_{1}=1.57$\text{\AA } and $d_{2}=1.54$\text{\AA }, showing that the four $sp^{3}$ hybridized bonds of carbon atoms are not totally equivalent, which is similar to h-diamond, but in contrast to c-diamond that has only single bond length $d$=$1.548$ \text{\AA }. Our calculations reveal that U$_n$-carbon (n=4-6) have similar mechanical properties with densities of $ 3.5 \sim 3.6$ g/cm$^3$, and Vickers hardness of $92.9 \sim 97.5$ GPa. The hardness of U$_n$-carbon was calculated by the method suggested in Ref. [\onlinecite{H}], and found to be comparable with or even larger than h- and c-diamond (Table I). The elastic constants, bulk modulus, Young's modulus and shear modulus
of U$_{n}$-carbon, c- and h-diamond were also calculated, which show that the elastic properties of U$_{n}$-carbon are in between those of c- and h-diamond (Table S1).

\begin{table*}[tbp]
\caption{Space group, lattice constants (a), equilibrium density ($\rho$), cohesive energy ($E_{c}$), energy gap ($E_{g}$),
 and Vickers hardness ($H_{v}$) \cite{H} of c-diamond, h-diamond, M-carbon and U$_n$-carbon (n=4-6)}%
\begin{tabular*}{18cm}{@{\extracolsep{\fill}}lccccccccc}
\hline\hline
Structure & Methond & Space group & a ({\AA }) & $\rho$ $(g/cm^{3})$& $E_{c}$ $(eV/atom)$ & $E_{g}$ $(eV)$ & $H_{\nu}$ $(GPa)$
\\ \hline
c-diamond & GGA (HSE06)& Fd$\bar{3}$m (OH$\bar{7}$)&a=b=c=3.565&3.490&7.643&4.32 $(5.59)$&92.59\\
c-diamond & LDA & Fd$\bar{3}$m (OH$\bar{7}$)&a=b=c=3.535&3.607&8.674 &4.41&94.86\\
c-diamond & Exp.\cite{diamondexp,diamondexp2}& Fd$\bar{3}$m (OH$\bar{7}$)&a=b=c=3.567& 3.520& 7.370& 5.45 & 96$\pm$5 \\
h-diamond & GGA (HSE06)&P63/mmc (D6H$\bar{4}$)&a=b=2.513,c=4.183&3.483&7.618&3.39 $(5.61)$&93.21\\
M-carbon  & GGA & C2/m (C2H$\bar{3}$)&a=b=4.769,c=4.15&3.332&7.48&3.69&90.43\\
U$_4$-carbon  &GGA (HSE06)&P63/mmc (D6H$\bar{4}$)&a=b=2.520,c=8.311&3.487&7.633&4.57 $(5.75)$&92.99\\
U$_4$-carbon  &LDA&P63/mmc (D6H$\bar{4}$)&a=b=2.492,c=8.222&3.605&8.664&4.58&95.07\\
U$_5$-carbon  &GGA (HSE06)&P$\bar{3}$M1 (D3D$\bar{3}$)&a=b=2.521,c=10.374&3.488&7.634&4.38 $(5.55)$&93\\
U$_5$-carbon  &LDA&P$\bar{3}$M1 (D3D$\bar{3}$)&a=b=2.494,c=10.263&3.605&8.665&4.40&93.05\\
U$_{6,I}$-carbon  &GGA (HSE06)&P63/mmc (D6H$\bar{4}$)&a=b=2.522,c=12.437&3.489&7.637&4.44 $(5.61)$&92.87\\
U$_{6,I}$-carbon  &LDA&P63/mmc (D6H$\bar{4}$)&a=b=2.494,c=12.304&3.605&8.668&4.46&94.94\\
U$_{6,II}$-carbon  &GGA (HSE06)&P$\bar{6}$m2 (D3H$\bar{1}$)&a=b=2.522,c=12.494&3.488&7.623&4.29 $(5.73)$&92.97\\
U$_{6,II}$-carbon  &LDA&P$\bar{6}$m2 (D3H$\bar{1}$)&a=b=2.490,c=12.362&3.600&8.652&4.00&97.53
 \\ \hline\hline
\end{tabular*}
\end{table*}

\begin{figure}[tbp]
\includegraphics[width=0.9\linewidth,clip]{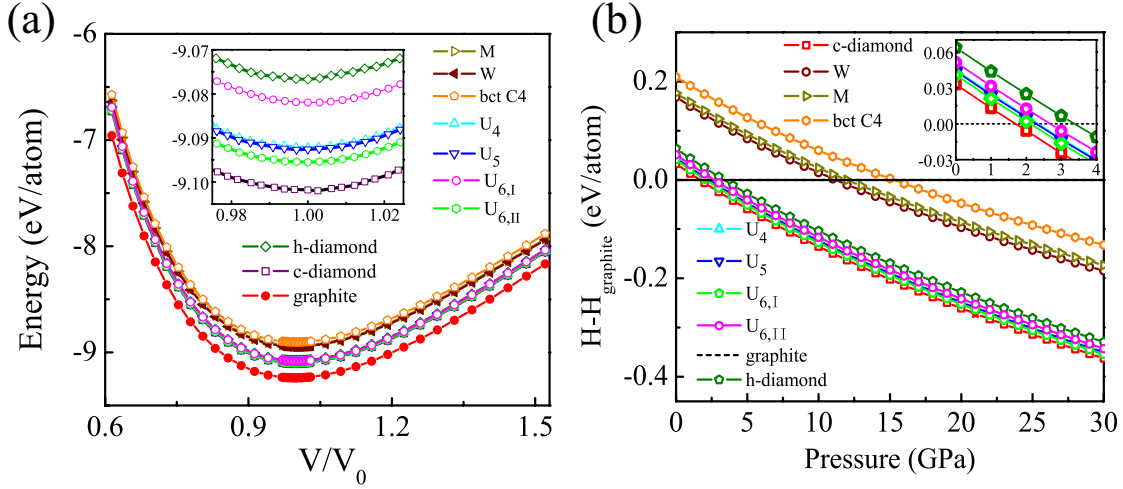}
\caption{(a) The total energy per atom versus volume $V$ for graphite, c-diamond, h-diamond, U$_4$-, U$_5$-, U$_{6,I}$-, U$_{6,II}$-, M-, W- and bct C4 carbon, where $V_{0}$ is the optimized volume. Inset is the enlarged part in the vicinity of $V/V_0=1$; (b) The enthalpy per atom versus pressure for c-diamond, h-diamond, U$_4$-, U$_5$-, U$_{6,I}$-, U$_{6,II}$-, M-, W- and bct C4 carbon with respect to graphite. Inset is the enlarged part at low pressures.}
\end{figure}

\bigskip

\noindent{\textbf{Thermodynamic and kinetic stabilities}}. To examine the stability of U$_n$-carbon, we calculated the total energy per atom versus volume $V$ for graphite, c-diamond, h-diamond, U$_n$- (n=4-6), M- \cite{Mcarbon}, W- \cite{wcarbon} and bct C4 \cite{bctc4} carbon for a comparison, where $V_{0}$ is the optimized lattice volume, as shown in Fig. 2(a). One may see that the total energy per atom versus the volume of U$_n$-carbon has a single minimum that corresponds to the optimized geometric structures, suggesting that U$_n$-carbon is energetically favorable. By comparing with other carbon allotropes, we observed that the total energy per atom of U$_n$-carbon is slightly higher than those of c-diamond and graphite, but is lower than those of h-diamond as well as M-, W- and bct C4 carbon, which implies that U$_n$-carbon are energetically more stable than h-diamond, M-, W- and bct C4 carbon, while they are metastable with respect to c-diamond and graphite. As h-diamond has been obtained experimentally, it is not impossible that U$_n$-carbon can be synthesized in laboratory or even already exist in nature. To explore the kinetic stability of U$_n$-carbon, the phonon spectra of U$_n$-carbon (n=4-6) were calculated, as presented in Fig. S3. No imaginary frequencies for these novel allotropes are found, indicating that U$_n$-carbon is also kinetically stable. Fig. 2(b) gives the enthalpy per atom with respect to that of graphite versus pressure for U$_n$-carbon (n=4-6). One may note that the curves are separated into two groups, where one group consists of those of M-, W-, and bct C4 carbon, and the other group is comprised of h-diamond, U$_n$-carbon and c-diamond. The enthalpies in the first group are dramatically higher than those of the other group. One may also note that the enthalpy of U$_n$-carbon is slightly higher than c-diamond but lower than that of h-diamond. This observation shows that U$_n$-carbon are thermodynamically more stable than M-, W-, bct C4 carbon as well as h-diamond.
\bigskip


\noindent{\textbf{Electronic structures}}. The electronic structures of U$_n$-carbon (n=4-6) are presented in Fig. 3. U$_{n}$-carbon
are insulators with indirect gaps, and their DOS are primarily contributed from $p$ electrons. Note that the tops of the valence band for U$_n$-carbon (n=4-6) all appear at $\Gamma$ point, while the bottoms of conducting band locate at other various points. For U$_4$-carbon, the bottom of the conducting band is between $\Gamma$ and $M$ point, leaving an indirect gap of 5.75 eV [Fig. 3(a)]; for U$_5$-carbon, the bottom of the conducting band is shifted to $K$ point, giving rise to an indirect gap of $5.55$ eV [Fig. 3(b)]; for U$_{6,I}$-carbon, the bottom of the conducting band sits between $\Gamma$ and $M$ points, giving the indirect gap of 5.61 eV [Fig. 3(c)]; while for U$_{6,II}$-carbon the bottom of the conducting band is at $H$ point, leading to the indirect gap of 5.73 eV [Fig. 3(d)]. In contrast to U$_{6,I}$-carbon, there is a clear downward of the conducting band between $H$ and $K$ points for U$_{6,II}$-carbon. \bigskip

\begin{figure}[tbp]
\includegraphics[width=0.8\linewidth,clip]{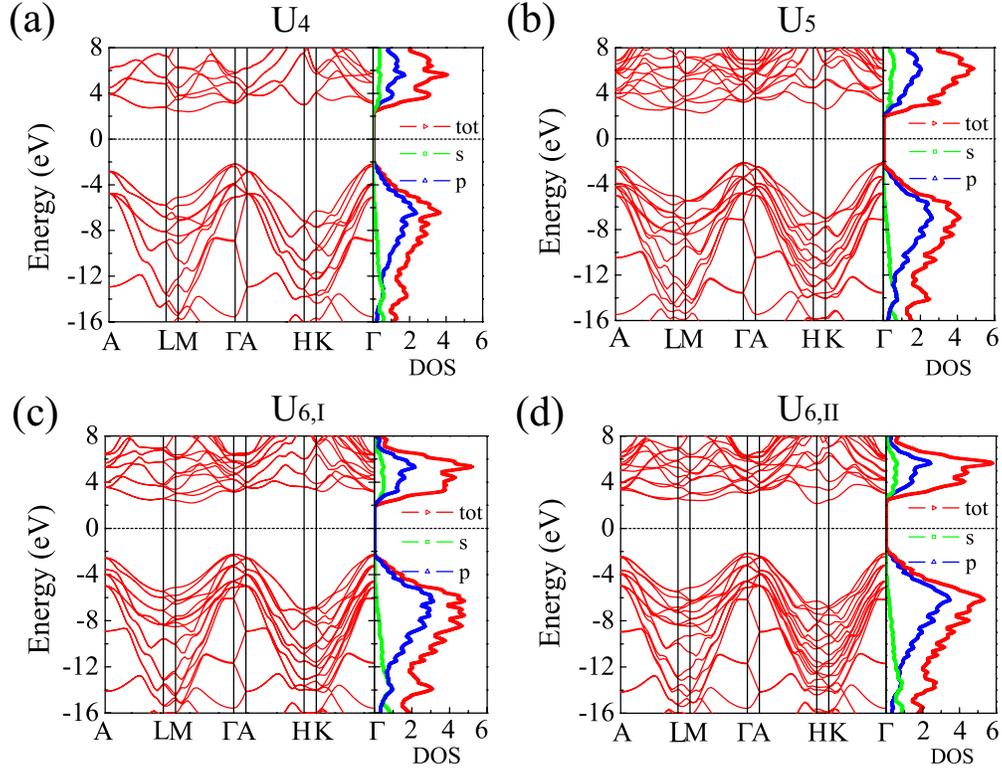}
\caption{The electronic band structures and density of states (DOS) of (a) U$_4$-carbon, (b) U$_5$-carbon, (c) U$_{6,I}$-carbon, and (d) U$_{6,II}$-carbon.}
\end{figure}

\noindent{\textbf{Simulated x-ray diffraction patterns and Raman spectroscopies}}. To provide more information useful for possible experimental realization, we simulated both x-ray diffraction (XRD) patterns and Raman spectroscopies of U$_{n}$-carbon (n=4-6), as presented in Fig. 4. The XRD spectra (at the wavelength 1.540562 \text{\AA}) are given in Fig. 4(a), where we also include the simulated XRD spectra of c- and h-diamond for a comparison. Our calculations reveal that the main peaks of c-diamond at 43.85$^\circ$, 75.1$^\circ$, 91.25$^\circ$ are related to crystal planes (111), (022) and (113), respectively, while those of h-diamond at 41.45$^\circ$, 43.2$^\circ$, 47$^\circ$, 75.6$^\circ$, 82.0$^\circ$ are related to crystal planes (010), (002), (011), (110)  and (013), respectively, which are in agreement with the experimental observations, showing the reliability of our present simulations. The main peaks of the XRD spectra of U$_n$-carbon (n=4-6) look apparently similar to those of h-diamond, but they have remarkable distinctions. For U$_4$-carbon, the peaks at 41.33$^\circ$, 42.8$^\circ$, 43.52$^\circ$, 46.98$^\circ$, 75.36$^\circ$, 82.39$^\circ$ and 91.27$^\circ$ correspond to the crystal planes (010), (011), (004), (012), (110), (016) and (114), respectively; for U$_5$-carbon, the peaks at 41.23$^\circ$, 42.24$^\circ$, 43.6$^\circ$, 44.98$^\circ$, 75.27$^\circ$, 77.84$^\circ$, and 91.24$^\circ$ correspond to the crystal planes (010), (011), (005), (012), (110), (017) and (015), respectively; for U$_{6,I}$-carbon, the peaks at 41.95$^\circ$, 43.6$^\circ$, 43.88$^\circ$, 46.97$^\circ$, 74.9$^\circ$, 75.36$^\circ$, 82.52$^\circ$
 and 91.26$^\circ$ correspond to the crystal planes (011), (006), (012), (013), (018), (110), (019) and (116), respectively; and for U$_{6,II}$-carbon, there is one more peak corresponding to the crystal planes (010) than U$_{6,I}$-carbon, suggesting that U$_{6,I}$- and U$_{6,II}$-carbon are really two different structures.

Raman spectroscopy can provide the unique fingerprint of intrinsic vibrational characteristics of a crystal.
The Raman modes and phonon frequencies of U$_{n}$-carbon (n=4-6) as well as c- and h-diamond are simulated, as presented in Fig. 4(b). The results show that for c-diamond there is only one three-fold degenerate Raman active mode $T_{2g}$ at $1332.8$ cm$^{-1}$, while for h-diamond there are three Raman active modes $A_{1g}$, $E_{1g}$, $E_{2g}$ at $1300.4$ cm$^{-1}$, $1331.8$ cm$^{-1}$, $1218.4$ cm$^{-1}$, respectively, whose mechanical representation is $\Gamma^{h-diamond}=A_{1g}+E_{2g}+E_{1g}$.
 These results coincide very well with earlier calculations and experimental observations \cite{ramanexp,ramanexp2,raman1,raman2,raman3}, revealing again the validity of our simulations. The Raman active modes of U$_{4}$-, U$_{5}$-, U$_{6,I}$- and U$_{6,II}$-carbon are found to be $\Gamma^{U_{4}}=2A_{1g}+2E_{2g}+2E_{1g}$, $\Gamma^{U_{5}}=5A_{1g}+5E_{g}$, $\Gamma^{U_{6,I}}=3A_{1g}+3E_{2g}+3E_{1g}$ and $\Gamma^{U_{6,II}}=6A'_{1}+5E'+6E''$, respectively. The Raman modes and corresponding phonon frequencies of U$_{n}$-carbon (n=4-6) are listed in Table S2 (supplementary information). The results show that U$_{n}$-carbon are novel $sp^3$ carbon allotropes that are different from the existing or previously predicted structures.
\bigskip

\begin{figure}[tbp]
\includegraphics[width=0.90\linewidth,clip]{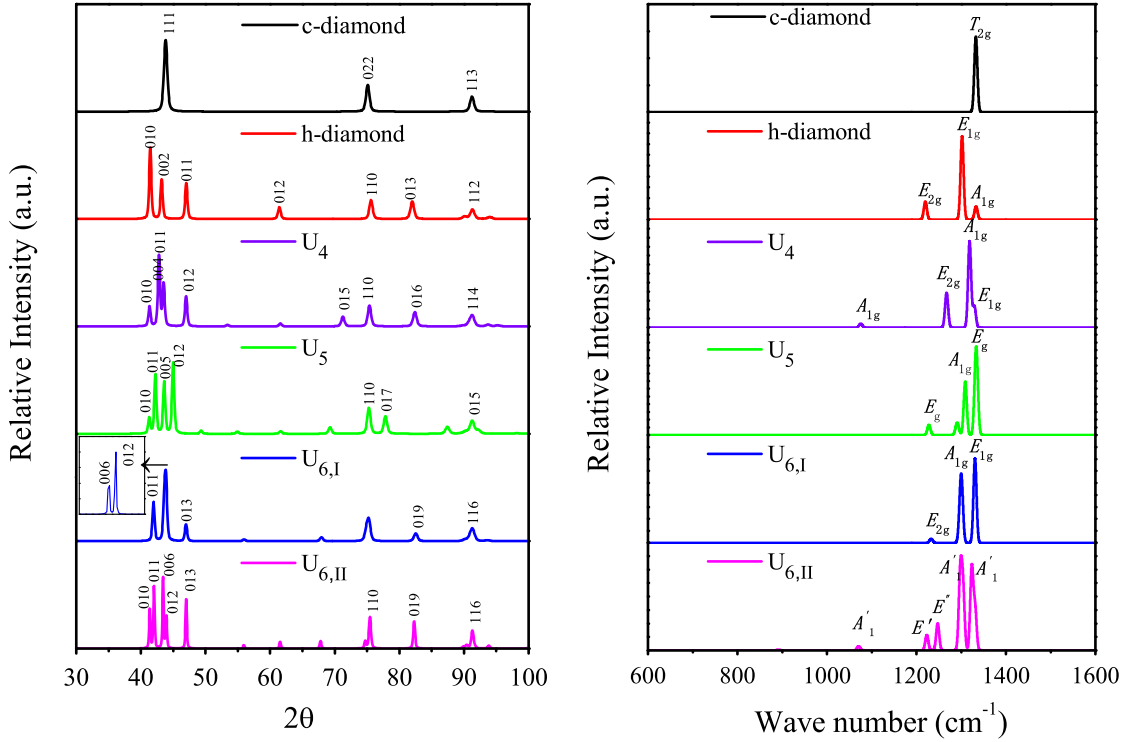}
\caption{(a) Simulated XRD patterns and (b) Raman spectra of c-diamond, h-diamond, U$_4$-, U$_5$-, U$_{6,I}$- and U$_{6,II}$-carbon. The wavelength of x-ray is presumed $d=1.54059$ \text{\AA }, and the Raman spectra were broadened by using the Gauss response shape with a full width at half maximum (FWHM) of 10 cm$^{-1}$.}
\end{figure}

\noindent{\textbf{Transforming pathways from graphite to U$_{n}$-carbon}}. In terms of the climbing image nudged elastic band method \cite{NEB}, we studied the transformation process from graphite to U$_{n}$-carbon. By utilizing the supercells of graphite as initial structures, we found that the properly structural arrangements of carbon atoms, as indicated in Figs. 5 (a)-(d), are of the most suitable from graphite to transform into U$_{n}$-carbon. For the whole transformation process, we calculated 33 steps, but for the purpose of illustration only 5 steps are presented in Fig. 5. For U$_{4}$-carbon, we take the supercell with 16 atoms of graphite that contains two layers of carbon atoms with zigzag buckling from the $x$ axis, as shown in Fig. 5(a). It is seen that the atoms are shifted along the $y$ axis with increasing the steps, and at step 28, a large portion of graphite are already transformed into $sp^{3}$ hybridized U$_{4}$-carbon, and at step 33, the graphite is totally transformed to U$_{4}$-carbon. For U$_{5}$-, U$_{6,I}$- and U$_{6,II}$-carbon, the initial structures of graphite are similar to that of U$_{4}$-carbon in the $x$ axis, but alter in the $z$ axis, where the supercells contain 40, 24 and 24 atoms, respectively. Since U$_{5}$-carbon has the closed packed period with odd number, to keep the translational invariance of graphite more atoms like 40 in a supercell should be taken. For a clarity of illustration, only half of the supercell is shown in Fig. 5(b). As they share similar structural symmetries, U$_{6,I}$- and U$_{6,II}$-carbon have the transformation processes similar to U$_{4}$-carbon, as revealed in Figs. 5(c) and 5(d). In addition, we also calculated the enthalpy versus transforming pathway steps from graphite to U$_n$-carbon at pressure 15 GPa [Fig. 5(e)]. One may observe that at initial transforming process, a relevant shift of two atom layers of graphite occurs, as also manifested in Figs. 5(a)-(d); and at step 19, phase transitions happen for U$_{n}$-carbon (n=4, 5, 6)  with energy barriers 0.280, 0.253, 0.243 and 0.241 eV, respectively. In addition, we also compared the energy barrier of U$_{n}$-carbon with those of c-diamond and h-diamond, and noted that the energy barriers of U$_{n}$-carbon are higher than c-diamond (0.226 eV) but lower than h-diamond (0.309 eV). When the step is over 24, the enthalpy of U$_{n}$-carbon (n=4, 5, 6) becomes negative, similar to the cases of c- and h-diamond, implying that in the course of formation the thermal energy was released, which also reflects that U$_{n}$-carbon (n=4,5,6) are more stable than graphite at high pressure. Therefore, it is reasonable to envision that U$_{n}$-carbon (n=4,5,6) can be obtained by compressing graphite at high temperature and high pressure.

\begin{figure}[tbp]
\includegraphics[width=0.65\linewidth,clip]{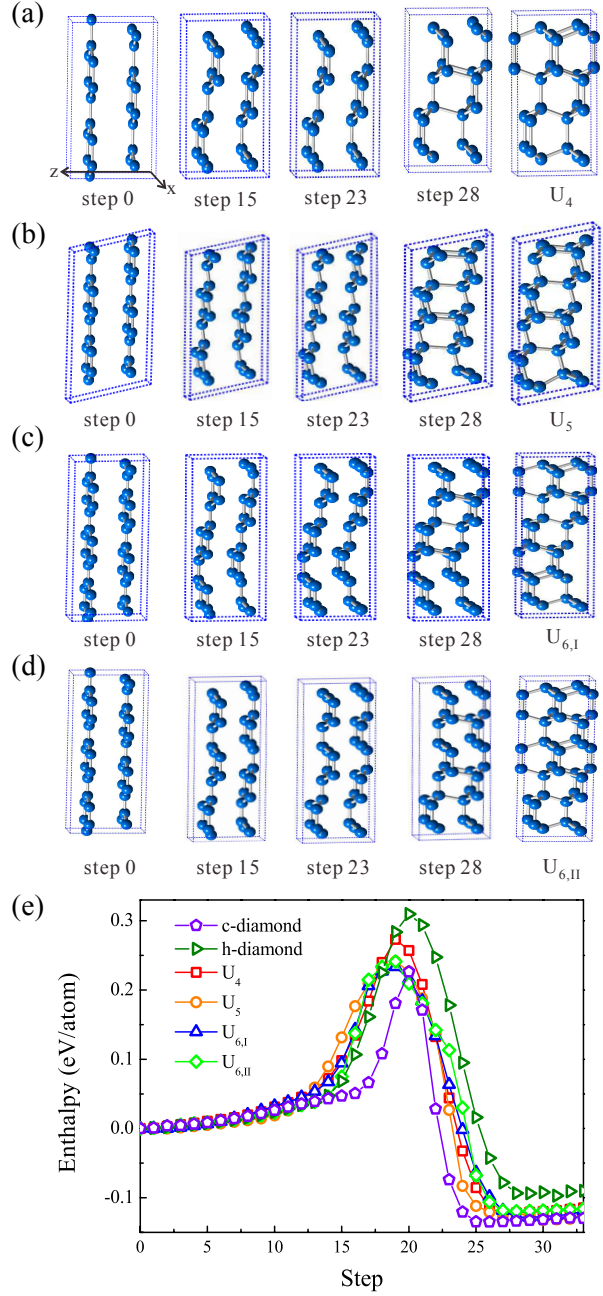}
\caption{The pathways from graphite to form U$_n$-carbon at pressure 15 GPa, where (a) U$_4$-carbon, (b) U$_5$-carbon, (c) U$_{6,I}$-carbon, and (d) U$_{6,II}$-carbon. (e) Enthalpy versus transforming pathway steps from graphite to U$_n$-carbon as well as c- and h-diamond at pressure 15 GPa.}
\end{figure}

\bigskip

\noindent{\textbf{Discussion}}

By means of the first-principles calculations we have found that there exists a family of new $sp^3$ carbon allotropes coined as U$_{n}$-carbon (n=2,3,4,5,$\cdots$), where h- and c-diamond are its simplest members with $n=2$ and $n=3$, respectively. The thermodynamic and kinetic stabilities of U$_{n}$-carbon were carefully examined and confirmed. It is shown that U$_{n}$-carbon are energetically even more favorable than h-diamond, and their cohesive energies are very close to those of c- and h-diamond, indicating that U$_{n}$-carbon can be probably obtained in laboratory or already exist in nature. The Vickers hardness of U$_{n}$-carbon is uncovered to be comparable or even harder than c- and h-diamond. The electronic structures reveal that U$_{n}$-carbon are insulators with indirect gaps. The simulated XRD pattern and Raman spectra demonstrate that U$_{n}$-carbon are quite different from other predicted or known carbon structures. In addition, we disclosed that U$_{6,I}$- and U$_{6,II}$-carbon at high pressure may be relevant to the recently debated new superhard phase of cold compression graphite (Fig. S4, supplementary information). Thus, U$_{n}$-carbon may be compelling superhard carbon allotropes that deserve to explore further, and would have potential applications in physics, chemistry and materials science.

\textit{Note added}: After completion of this present paper, we were informed that nanopolycrystal diamond with the structures similar to U$_{n}$-carbon was synthesized by direct-conversion method from graphite \cite{tanigaki}.
\bigskip

\noindent{\textbf{Methods}}

The structural relaxation, mechanical and electronic properties of U$_n$-carbon were calculated
using the Vienna \textit{ab initio} simulation package (VASP) \cite{vasp1,vasp2} with
the projector augmented wave (PAW) method \cite{paw}. Both local density approximation
(LDA) developed by Perdew and Zunger \cite{ldapz} and generalized gradient approximation (GGA)
 developed by Perdew and Wang \cite{ggapw} were adopted for exchange-correlation potentials.
The Heyd-Scuseria-Ernzerhof (HSE06) \cite{HSE06} screened Coulomb hybrid density functional was also used.
 The total energy was converged to within 1 $meV$ with the plane-wave cutoff energy 520 $eV$.
 A mesh of $11\times
11\times 11$ $k$-point in Monkhorst-Pack scheme \cite{MPscheme} was used to sample the Brillouin zone. The geometries were optimized until the remanent Hellmann-Feynman forces on the ions were less than 0.01 $eV/nm$. The calculations of phonon and Raman spectra \cite{Raman} were performed by Quantum-ESPRESSO package \cite{espresso}.

\bigskip

\noindent{\textbf{Acknowledgements}}

All calculations were completed in the Supercomputing Center of CAS (Shenteng 7000) and Shanghai Supercomputer Center (MagicCube).
This work is supported in part by the NSFC (Grants No. 90922033, No. 10934008, No.
10974253 and No. 11004239), the MOST of China (Grant No.
2012CB932900, 2013CB933401), and the CAS.
\bigskip

\noindent{\textbf{Author contributions}}

HJC performed all \textit{ab initio} calculations. HJC and GS wrote the paper. All authors discussed the results and reviewed the paper.
\bigskip

\noindent{\textbf{Additional information}}

Supplementary Information accompanies this paper. The authors declare no competing financial interests.
\bigskip

\noindent{\textbf{Figure Captions}} \bigskip

\noindent{\textbf{Figure 1} The geometrical structures of novel carbon allotropes U$_n$-carbon in a unit cell. (a) U$_4$-carbon ($n=4$, ABCB), (b) U$_5$-carbon ($n=5$, ACBAC), (c) U$_{6,I}$-carbon  ($n=6$, ABACBC), and (d) U$_{6,II}$-carbon ($n=6$, ABABAC). The letters A, B and C indicate the inequivalent positions of C-C unit, which also represent the inequivalent sites in the closed-packed planes from top view.} \\

\noindent{\textbf{Figure 2} (a) The total energy per atom versus volume $V$ for graphite, c-diamond, h-diamond, U$_4$-, U$_5$-, U$_{6,I}$-, U$_{6,II}$-, M-, W- and bct C4 carbon, where $V_{0}$ is the optimized volume. Inset is the enlarged part in the vicinity of $V/V_0=1$; (b) The enthalpy per atom versus pressure for c-diamond, h-diamond, U$_4$-, U$_5$-, U$_{6,I}$-, U$_{6,II}$-, M-, W- and bct C4 carbon with respect to graphite. Inset is the enlarged part at low pressures.} \\

\noindent{\textbf{Figure 3} The electronic band structures and density of states (DOS) of (a) U$_4$-carbon, (b) U$_5$-carbon, (c) U$_{6,I}$-carbon, and (d) U$_{6,II}$-carbon.} \\

\noindent{\textbf{Figure 4} (a) Simulated XRD patterns and (b) Raman spectra of c-diamond, h-diamond, U$_4$-, U$_5$-, U$_{6,I}$- and U$_{6,II}$-carbon. The wavelength of x-ray is presumed $d=1.54059$ \text{\AA }, and the Raman spectra were broadened by using the Gauss response shape with a full width at half maximum (FWHM) of 10 cm$^{-1}$.} \\

\noindent{\textbf{Figure 5} The pathways from graphite to form U$_n$-carbon at pressure 15 GPa, where (a) U$_4$-carbon, (b) U$_5$-carbon, (c) U$_{6,I}$-carbon, and (d) U$_{6,II}$-carbon. (e) Enthalpy versus transforming pathway steps from graphite to U$_n$-carbon as well as c- and h-diamond at pressure 15 GPa.} \\

\end{document}